\documentclass[journal]{IEEEtran}

\ifCLASSINFOpdf
\else
   \usepackage[dvips]{graphicx}
\fi
\usepackage{url}

\usepackage{graphicx}
\usepackage{amsmath}
\usepackage{bm}
\usepackage{amsopn}
\usepackage{multirow}
\usepackage{booktabs}
\usepackage{amssymb}
\usepackage{algorithm}  
\usepackage{algorithmicx}  
\usepackage{amsmath}
\usepackage{algpseudocode}
\usepackage{cite}

\begin{document}

\title{A Novel Blind Source Separation Framework Towards Maximum Signal-To-Interference Ratio}

\author{Jianjun Gu, Longbiao Cheng, Dingding Yao, Junfeng Li, and Yonghong Yan
\thanks{This research is partially supported by the National Key Research and Development Program of China (No. 2020YFC2004104) and the National Natural Science Foundation of China (Nos. 12104483, 11911540067).}
\thanks{All the authors are with the Key Laboratory of Speech Acoustics and Content Understanding, Institute of Acoustics, Chinese Academy of Sciences, and University of Chinese Academy of Sciences, Beijing 100190, China. (e-mail: junfeng.li.1979@gmail.com)}}

\markboth{Journal of \LaTeX\ Class Files, Vol. 14, No. 8, August 2015}
{Shell \MakeLowercase{\textit{et al.}}: Bare Demo of IEEEtran.cls for IEEE Journals}
\maketitle

\begin{abstract}
This letter proposes a new blind source separation (BSS) framework termed minimum variance independent component analysis (MVICA), which can potentially achieve the maximum output signal-to-interference ratio (SIR) while also allowing more flexibility in real implementations. The statistical independence assumption has been the foundation of the most dominant BSS techniques in recent decades. However, this assumption does not always hold true and the accurate probabilistic modeling of source is inherently difficult. To overcome these limitations and improve the separation performance, the MVICA framework is rigorously derived by optimizing the design of these independence-based BSS algorithms with the maximum SIR criterion. A deep neural network-supported implementation of MVICA is subsequently described. Experimental results under various conditions show the superiority of MVICA over the state-of-the-art BSS algorithms, in terms of not only SIR but also signal-to-distortion ratio and automatic speech recognition rate.
\end{abstract}


\IEEEpeerreviewmaketitle

\section{Introduction}
\label{sec:intro}
\IEEEPARstart{B}{lind} source separation (BSS) technology~\cite{BSS} aims at determining the original source signals from the observed mixtures with minimal prior knowledge about the mixing process, which has found prominent applications in speech communication and speech-based human-machine interaction systems~\cite{BSS2}.
For BSS in the (over)determined case (number of microphones $\geq$ number of sources), statistical independence between the source signals is widely assumed to estimate the demixing matrix \cite{independence1,independence2}, bringing out a family of independence-based BSS algorithms. Among them, frequency domain independent component analysis (FDICA) \cite{FDICA1,FDICA2} and independent vector analysis (IVA) \cite{IVA1,IVA2,IVA3} are the most classical ones.
The auxiliary function technique is then introduced to more stably and efficiently update the demixing matrices for ICA (AuxICA) \cite{AuxICA} and IVA (AuxIVA) ~\cite{AuxIVA1,AuxIVA2}.
The separation performance of these methods highly relies on the quality of the source model, which thus attracts wide research interests.
The independent low-rank matrix analysis (ILRMA) algorithm \cite{ILRMA} incorporate a nonnegative matrix factorization (NMF) decomposition-based source model into IVA, leading to significant improvements in music source separation.
Given the powerful modeling capability of deep neural network (DNN), DNN is recently utilized to model the spectral pattern of signals and achieves the state-of-the-art (SOTA) BSS performance \cite{IDLMA,IDLMA1,DIVA_Drude,DIVA_KANG,DIVA_ROBIN}.

Despite the foregoing improvements, there is still a problem that the maximum independence criterion does not always lead to a satisfactory separation result. Because the independence assumption does not always hold true \cite{limits_BSS,independence_test} and the precise modeling of source remains a challenging problem \cite{DIVA_KANG}. 
As a supplement to the maximum independence criterion, additional penalties such as the geometrical constraints \cite{DIVA1,DIVA2,DIVA3} and the sparsity-based constraints \cite{SIVA,HVA}, have been demonstrated to improve the performance when incorporated into these BSS methods.
However, these penalties rely on prior knowledge of sources \cite{HVA}. 
By contrast, the maximum signal-to-interference ratio (SIR) criterion \cite{GEV} directly optimizes the separation performance with respect to interference suppression and target source preservation, and can be generally used for any sources.
Unfortunately, how to incorporate the maximum SIR criterion into these BSS methods remains a problem.

In this letter, our goal is to maximize the output SIR for each source while preserving the advantage of fast and stable solutions from the mainstream independence-based BSS algorithms.
To this end, we follows the auxiliary function technique-based update rules \cite{AuxICA,AuxIVA2} in these algorithms, but propose to optimize the choice of the weighted covariance matrix toward maximizing the output SIR.
The optimal weighted covariance matrix is proved to be the interference covariance matrix, form which the maximum-SIR demixing matrix can be directly computed, thus also avoiding the difficulty of source modeling.
As a result, a new BSS framework called minimum variance independent component analysis (MVICA) is obtained.
Any methods for estimating the interference covariance matrix can be incorporated into MVICA, therefore giving more freedom to actual applications and further studies.
A DNN-supported implementation of MVICA is finally presented. Experiments reveal that MVICA significantly outperforms the SOTA BSS algorithms, in terms of SIR, signal-to-distortion ratio (SDR) and automatic speech recognition (ASR) rate.

\section{Problem Formulation}
We address the determined BSS problem where $K$ sources are observed by $K$ microphones ($K\geq2$). The short-time Fourier Transform (STFT) coefficients of the source, the observed, and the estimated signals are denoted as $s_k(f,t)$, $x_k(f,t)$ and $y_k(f,t)$, respectively, where $t=1,...,N; f=1,...,N_f;$ and $k=1,...,K$ index the time frames, frequency bins and sources, respectively. 
For the reverberant mixtures, 
when the window size used in the STFT is sufficiently long compared with the impulse responses between sources and microphones, 
the mixing process can be modeled by
\begin{equation}
	\label{mix_model}
	\bm{x}(f,t)={A}(f)\bm{s}(f,t),
\end{equation}
where $\bm{d}(f,t)=[d_1(f,t), d_2(f,t), \cdots, d_K(f,t)]^T \in \mathbb{C}^{K} (d \in \{s,x,y\})$ denotes the vector representation of signal, and
\begin{equation}
	{A}(f)=[\bm{a}_1(f),\bm{a}_2(f), \cdots, \bm{a}_K(f)] \in \mathbb{C}^{K \times K},
\end{equation}
is the mixing matrix containing the acoustic transfer functions from the source positions to the microphones, and the superscript $^T$ denotes the transpose.

The primary aim of BSS is to estimate a demixing matrix
\begin{equation}
	W(f)=[\bm{w}_1(f),\bm{w}_2(f), \cdots, \bm{w}_K(f)]^H \in \mathbb{C}^{K \times K},
\end{equation}
so that the source signals can be recovered by
\begin{equation}
	\label{Y-W-A-X}
	\bm{y}(f,t)={W}(f)\bm{x}(f,t),
\end{equation}
and the superscript $^H$ denotes the Hermitian transpose. 

\section{Conventional Method}
The conventional independence-based BSS algorithms normally estimate the demixing matrix $W(f)$ as the one that maximizes measure related to independence of estimated sources.
Let $p\left(\bm{y}_k(t)\right)$ be the probabilistic model, many of them fall into a optimization problem of the following form \cite{BSS_review}:
\begin{equation}
	\label{obj_function}
	\mathop{\arg\min}_{\left[W(f)\right]_{f=1}^{N_f}} \sum_{k,t} G\left\{\bm{y}_k(t)\right\} -\sum_{f} \log |\operatorname{det} W(f)|,
\end{equation}
where $G\left\{\bm{y}_k(t)\right\}=-\log p\left(\bm{y}_k(t)\right)$ is the contrast function, $\operatorname{det}(\cdot)$ denote the determinant operator and
\begin{equation}
	\bm{y}_k(t)=[y_k(1,t),...,y_k(N_f,t)]^T \in \mathbb{C}^{N_f}.
\end{equation}
For example, FDICA, IVA and ILRMA can be obtained by assuming the following source models \cite{BSS_review}, respectively:
\begin{equation}
	\begin{aligned}
		\text{FDICA} &\rightarrow p\left(\bm{y}_k(t)\right) \propto \prod_f \exp\left\{-\left|y_k(f,t)\right|\right\},\\
		\text{IVA} &\rightarrow p\left(\bm{y}_k(t)\right) \propto \exp\left\{-\left\|\bm{y}_k(f,t)\right\|_2 \right\},\\
		\text{ILRMA} &\rightarrow p\left(\bm{y}_k(t)\right) \propto \prod_f \exp\left\{-\frac{\left|y_k(f,t)\right|^2} {\tilde{y}_k(f,t)}\right\},
	\end{aligned}
\end{equation}
where $\tilde{y}_k(f,t)$ denotes the source variance estimated by NMF and $\|\cdot\|_2$ denotes $L_2$ norm of a vector.

To solve Eq.(\ref{obj_function}) based on the auxiliary function technique \cite{AuxIVA2}, the weighted covariance matrix is further introduced as
\begin{equation}
	\label{V}
	V_{k}(f)=E\left \{ \frac{G^{'}\left(\bm{y}_{k}(t)\right)}{r_k(f,t)} \bm{x}(f,t)\bm{x}^{H}(f,t) \right \},
\end{equation}
where $G^{'}$ denotes the differential of $G$, $E\{\cdot\}$ denote expectation over frames and $r_k(f,t)$ gets assigned the value of $|y_k(f,t)|$, $\left\|\bm{y}_k(f,t)\right\|_2$ and $|\tilde{y}_k(f,t)|$ for FDICA, IVA and ILRMA, respectively.
Fast and stable demixing matrix update rules can then be obtained as
\begin{equation}
	\label{demxing_matrix_estimation_1}
	\bm{w}_{k}(f) \leftarrow V_{k}^{-1}(f) W^{-1}(f)\bm{e}_k,
\end{equation}
\begin{equation}
	\label{demxing_matrix_estimation2}
	\bm{w}_{k}(f) \leftarrow \bm{w}_{k}(f)/\sqrt{\bm{w}^{H}_{k}(f)V_{k}(f)\bm{w}_{k}(f)},
\end{equation}
where $\bm{e}_k$ denotes the unit vector with the $k$th element unity. 
In practice, Eqs.(\ref{Y-W-A-X}), (\ref{V}), (\ref{demxing_matrix_estimation_1}) and (\ref{demxing_matrix_estimation2}) are applied in order for all $k$ and $f$. The iteration is repeated until convergence is obtained.

\section{Proposed Method}
\subsection{Derivation of The MVICA Algorithm Framework}
Our goal is to determine the maximum-SIR demixing matrix with the above update rules, 
where $\bm{w}_{k}(f)$ can be directly estimated from $V_k(f)$ by substituting Eq.(\ref{demxing_matrix_estimation_1}) into (\ref{demxing_matrix_estimation2}) as
\begin{equation}
	\label{demixingmatrix_V}
	\bm{w}_{k}(f) \leftarrow \frac{V_{k}^{-1}(f)\tilde{\bm{a}}_k(f)}{\sqrt{\tilde{\bm{a}}_k(f)V_{k}^{-1}(f))\tilde{\bm{a}}_k(f)}},
\end{equation}
\begin{equation}
	\label{estmated_a}
	\tilde{\bm{a}}_k(f) =\tilde{W}^{-1}(f)\bm{e}_k.
\end{equation}
$\tilde{W}(f)$ in Eq.(\ref{estmated_a}) is the estimated demixing matrix obtained in the previous iteration. 
So we first determine the optimal $V_k(f)$,
from which the maximum-SIR demixing matrix can then be computed using Eqs.(\ref{demxing_matrix_estimation_1}) and (\ref{demxing_matrix_estimation2}). $V_k(f)$ depends on the source models as indicated in Eq.(\ref{V}), but we avoid the difficulty of source modeling and directly derive its optimal form, since a explicit source model is not always necessary \cite{HVA}.
The optimization problem is thus formulated as
\begin{equation}
	\label{V_Optimize}
	{V}_k^{\text{opt}}(f)=\mathop{\arg\max}\limits_{V_k(f)} \text{SIR}_k(f)
\end{equation}
where the output SIR is defined as
\begin{equation}
	\label{SIR}
	 \text{SIR}_k(f) = \frac{\bm{w}_{k}^{H}(f)\bm{\Phi}_{\bm{S}_k}(f)\bm{w}_{k}(f)}{\bm{w}_{k}^{H}(f)\bm{\Phi}_{\bm{N}_k}(f)\bm{w}_{k}(f)}
\end{equation}
and
$\bm{\Phi}_{\bm{S}_k}(f)$ and $\bm{\Phi}_{\bm{N}_k}(f)$ denote the covariance matrix of the $k$th source image and the corresponding interference, respectively, i.e.,
\begin{equation}
	\label{source_cov}
	\bm{\Phi}_{\bm{S}_k}(f) =  E\left\{\bm{S}_k(f,t)\bm{S}_k^{H}(f,t)\right\} \in \mathbb{C}^{K \times K},
\end{equation}
\begin{equation}
	\label{interference_cov}
	\bm{\Phi}_{\bm{N}_k}(f) =  E\left\{\bm{N}_k(f,t)\bm{N}_k^{H}(f,t)\right\} + \mathcal{E} * I \in \mathbb{C}^{K \times K},
\end{equation}
\begin{equation}
	\label{reverb_source}
	\bm{S}_k(f,t)= \bm{a}_k(f)s_k(f,t),
\end{equation}
\begin{equation}
	\label{interference}
	\bm{N}_k(f,t)=\bm{x}(f,t)-\bm{S}_k(f,t),
\end{equation}
where a diagonal loading term $\mathcal{E} * I$ ($\mathcal{E}$ is a tiny constant and $I$ denotes the identity matrix) is introduced into 
$\bm{\Phi}_{\bm{N}_k}(f)$ to stabilize the matrix inverse operation, it is thus positive definite and can decomposed as
\begin{equation}
	\label{interference_cov2}
	\bm{\Phi}_{\bm{N}_k}(f)=\bm{P}^H_k(f)\bm{P}_k(f),
\end{equation}
where $\bm{P}_k(f)$ is the $K \times K$ invertible matrix. Substituting Eq.(\ref{reverb_source}) into (\ref{source_cov}), $\bm{\Phi}_{\bm{S}_k}(f)$ is a rank-1 matrix given by 
\begin{equation}
	\label{source_cov2}
	\bm{\Phi}_{\bm{S}_k}(f) = E\left\{\left|s_k(f,t)\right|^2\right\}\bm{a}_k(f)\bm{a}_k^{H}(f).
\end{equation}

Substituting Eqs.(\ref{interference_cov2}), (\ref{source_cov2}) and (\ref{demixingmatrix_V}) into (\ref{V_Optimize}), the optimization problem can then be rewritten as
\begin{equation}
	\label{V_Optimize2}
	\mathop{\arg\max}\limits_{V_k(f)} \frac{E\left\{\left|s_k(f,t)\right|^2\right\}\bm{D}^H\bm{P}^{-H}_k(f)\bm{a}_k(f)\bm{a}_k^{H}(f)\bm{P}^{-1}_k(f)\bm{D}}{\bm{D}^H\bm{D}},
\end{equation}
where
\begin{equation}
	\bm{D}=\bm{P}_k(f)V_{k}^{-1}(f)\tilde{\bm{a}}_k(f).
\end{equation}
According to the Cauchy-Schwarz inequality \cite{cauchy_inequa}, we have 
\begin{equation}
	\begin{aligned}
		\bm{D}^H&\bm{P}^{-H}_k(f)\bm{a}_k(f)\bm{a}_k^{H}(f)\bm{P}^{-1}_k(f)\bm{D}  \\ \leq &\left\{\bm{D}^H\bm{D}\right\}\left\{\bm{a}_k^{H}(f)\bm{P}^{-1}_k(f)\bm{P}^{-H}_k(f)\bm{a}_k(f)\right\}.
	\end{aligned}
\end{equation}
Thus, the output SIR satisfies
\begin{equation}
	\label{maximum_SIR}
	\text{SIR}_k(f) \leq E\left\{\left|s_k(f,t)\right|^2\right\} \bm{a}_k^H(f)\bm{\Phi}_{\bm{N}_k}^{-1}(f)\bm{a}_k(f)
\end{equation}
and the equality sign is satisfied if and only if
\begin{equation}
	\lambda  \bm{P}^{-H}_k(f)\bm{a}_k(f) =  \bm{D},
\end{equation}
that is,
\begin{equation}
	\lambda  \bm{\Phi}_{\bm{N}_k}^{-1}(f)\bm{a}_k(f)=V_{k}^{-1}(f)\tilde{\bm{a}}_k(f),
\end{equation}
where $\lambda$ is an positive constant, and $\tilde{\bm{a}}_k(f)$ defined in Eq.(\ref{estmated_a}) is the approximation of $\bm{a}_k(f)$ since the desired demixing matrix should satisfy $\tilde{W}(f) \approx A^{-1}(f)$. 
Therefore, the weighted covariance matrix that is capable of the maximum SIR is exactly the interference covariance matrix, i.e.,
\begin{equation}
	\label{optimal_V}
	{V}_k^{\text{opt}}(f) = \frac{1}{\lambda} \bm{\Phi}_{\bm{N}_k}(f),
\end{equation}
where the scale parameter $1/\lambda$ can be omitted because of the normalization operation followed in Eq.(\ref{demxing_matrix_estimation2}).

By replacing Eq.(\ref{V}) with Eq.(\ref{optimal_V}) for the conventional independence-based BSS algorithm,  we obtain the MVICA algorithm framework.
As summarized in \textbf{Algorithm 1}, MVICA is compose of two discrete steps: estimating the interference covariance matrix for each source, which in turn is used to blindly compute the demixing matrix $\bm{w}_{k}(f)$ based on the update rules given in Eq.(\ref{demxing_matrix_estimation_1}).
It is worth noting that MVICA omits the normalization operation in Eq.(\ref{demxing_matrix_estimation2}).
This computational simplification comes because the scale ambiguity for the separated signals will be resolved by the post processing. 
 
\subsection{Differences Between MVICA And The Existing Methods}
As given in Eq.(\ref{maximum_SIR}), the maximum SIR obtained by MCIVA is equivalent to that of the minimum variance distortionless response (MVDR) beamformer \cite{MVDR,MVDR2,sen_analy} and the generalized eigenvalue (GEV) beamformer \cite{GEV}.
Moreover, both MVICA and the GEV beamformer strive to maximize the out SIR.
However, different from the update rules adopted by MVICA, the GEV beamformer and its extension in \cite{IVE} estimate the filter coefficients via GEV decomposition, and we will compare their performances in the experiment part.
MVICA is also preferred over the MVDR beamformer and its DNN-supported variants \cite{neuralbeamformer1,neuralbeamformer2,neuralbeamformer3}, because it avoids the extra estimation of the relative transfer function (RTF) \cite{neuralbeamformer1} or the covariance matrix of sources \cite{neuralbeamformer3}, which is a remarkable benefit.

\subsection{DNN-based Implementation of MVICA}
The key problem now is the accurate estimation of the interference covariance matrix $\bm{\Phi}_{\bm{N}_k}(f)$.
Any methods for estimating $\bm{\Phi}_{\bm{N}_k}(f)$ can be incorporated into MVICA.
Inspired by success of the mask-based beamformers \cite{neuralbeamformer1,neuralbeamformer2,neuralbeamformer3},
we introduce a DNN-supported implementation of MVICA, which estimates the interference signal by
\begin{equation}
	\tilde{\bm{N}}_k(f,t)=\bm{m}_k(f,t)\bm{x}(f,t),
\end{equation}
where $\bm{m}_k \in \mathbb{C}^{N_f \times T}$ is the complex-valued mask computed by the DNN and its weights are shared across channels. The interference covariance matrix can then be computed as:
\begin{equation}
	\tilde{\bm{\Phi}}_{\bm{N}_k}(f) = \frac{\sum_{t=1}^{T}\tilde{\bm{N}}_k(f,t)\tilde{\bm{N}}_k^H(f,t)}{\sum_{t=1}^{T}\bm{m}_k^H(f,t)\bm{m}_k(f,t)} + \mathcal{E} * I.
\end{equation}
\begin{algorithm}[!htbp]
	\caption{The MVICA algorithm framework}
	\begin{algorithmic}
		\State 1. Estimate the interference covariance matrix $\tilde{\bm{\Phi}}_{\bm{N}_k}(f)$ for all $k$.
		\State 2. Compute the demixing matrix for all $f$:
		\For{$l$ of number of iterstions $L$}
		\For{$k=1,2,...,K$}
		\begin{equation}
			\bm{w}_{k}(f) \leftarrow \tilde{\bm{\Phi}}_{\bm{N}_k}^{-1}(f) W^{-1}(f)\bm{e}_k.
		\end{equation}
		\EndFor
		\EndFor
	\end{algorithmic}
\end{algorithm}

For the DNN architecture, the deep complex convolutional recurrent network (DCCRN) \cite{DCCRN} is adopted given its outstanding performance in speech enhancement.
The multi-channel complex mixture spectrograms are stacked as input features to estimate $\bm{m}_k$ for all $k$ simultaneously.
For the loss function, we use the complex-domain scale-invariant signal-to-distortion ratio in \cite{WPD++} combined with the spectral mean square error loss to measure the estimation error of $\tilde{\bm{N}}_k(f,t)$. Permutation invariant training \cite{UPIT} is also utilized because of the ambiguity of the output order. 
Unlike the previous DNN-supported methods \cite{IDLMA,IDLMA1,DIVA_Drude,DIVA_ROBIN}, which update the source model parameters with DNN iteratively, leading to the expensive computational cost and the mismatch between training and testing data, the proposed MVICA algorithm learn the mask directly from the mixtures signals.

\section{Experimental Results}
\begin{table*}[!htb]
	\caption{Separation performances of the different methods on the test mixing data ($K$: the number of sources).}	
	\centering
	\begin{tabular}{cccccccc|ccccccc|cc}
		\hline \hline \specialrule{0pt}{1pt}{1pt}
		\multirow{2}{*}{$\bm{K}$} &  & \multirow{2}{*}{\textbf{Methods}} &  & \multicolumn{4}{c|}{\textbf{SIR Improvements (dB)}}                        & \multicolumn{4}{c}{\textbf{SDR Improvements (dB)}}                       &  & \multicolumn{2}{c|}{\textbf{WER(\%)}}    & \multicolumn{2}{c}{\textbf{CER(\%)}}    \\ \specialrule{0pt}{1pt}{1pt} \cline{5-12} \cline{14-17} \specialrule{0pt}{1pt}{1pt}
		&  &                          &  & 100ms          & 200ms          & 300ms          & 400ms          & 100ms          & 200ms          & 300ms         & 400ms         &  & 100ms          & 200ms          & 100ms         & 200ms          \\ \specialrule{0pt}{1pt}{1pt} \hline \hline \specialrule{0pt}{1pt}{1pt}
		\multirow{5}{*}{2} &  & AuxIVA \cite{AuxIVA2}                   &  & 11.28          & 9.91           & 8.71           & 7.56           & 7.44           & 5.51           & 3.95          & 2.90          &  & 48.56          & 60.52          & 32.31         & 39.04          \\
		&  & ILRMA \cite{ILRMA}                    &  & 16.54          & 12.03          & 10.50          & 8.83           & 11.90          & 7.02           & 4.79          & 3.27          &  & 39.53          & 59.08          & 24.90         & 39.73          \\
		&  & Kang \cite{DIVA_KANG}                    &  & 21.46          & 15.72          & 13.90          & 11.44          & 15.58          & 9.51           & 6.69          & 4.65          &  & 17.03          & 25.95          & 9.20          & 14.61          \\
		&  & IDLMA \cite{IDLMA}                   &  & 20.90          & 16.14          & 14.04          & 11.89          & 16.19          & 10.13          & 6.94          & 5.01          &  & 20.83          & 42.46          & 12.89         & 27.26          \\
		&  & MVICA-DNN                    &  & \textbf{25.92} & \textbf{20.54} & \textbf{17.70} & \textbf{15.22} & \textbf{18.65} & \textbf{11.63} & \textbf{7.52} & \textbf{5.61} &  & \textbf{8.36}  & \textbf{16.97} & \textbf{3.77} & \textbf{7.70}  \\ \specialrule{0pt}{1pt}{1pt} \hline \specialrule{0pt}{1pt}{1pt}
		\multirow{5}{*}{3} &  & AuxIVA \cite{AuxIVA2}                  &  & 9.12           & 7.13           & 6.45           & 5.46           & 5.72           & 3.95           & 3.13          & 2.21          &  & 93.57          & 115.60         & 52.78         & 70.19          \\
		&  & ILRMA \cite{ILRMA}                   &  & 12.02          & 8.57           & 7.03           & 5.41           & 8.56           & 5.14           & 3.53          & 2.08          &  & 76.36          & 112.07         & 41.19         & 66.88          \\
		&  & Kang \cite{DIVA_KANG}                    &  & 13.59          & 11.29          & 9.32           & 10.11          & 9.83           & 7.14           & 5.15          & 5.20          &  & 51.39          & 64.42          & 25.48         & 34.98          \\
		&  & IDLMA \cite{IDLMA}                   &  & 17.87          & 13.51          & 9.99           & 12.26          & 13.10          & 9.06           & 5.85          & 6.57          &  & 23.83          & 36.61          & 11.21         & 20.18          \\
		&  & MVICA-DNN                    &  & \textbf{19.16} & \textbf{16.39} & \textbf{13.73} & \textbf{13.74} & \textbf{13.36} & \textbf{10.37} & \textbf{7.60} & \textbf{6.80} &  & \textbf{16.47} & \textbf{32.58} & \textbf{7.26} & \textbf{16.70} \\ \specialrule{0pt}{1pt}{1pt} \hline
	\end{tabular}
\end{table*}
\subsection{Data Preparation and Experimental Setup}
We train our models on synthesized convolutive mixtures of speech. 14,400 speech samples are selected from the WSJ0 dataset \cite{wsj0} in total. While, 150 room impulse responses (RIRs) are generated for each of the 2-source and 3-source case using the image source method \cite{image_method} in a simulated 4 m $\times$ 4 m $\times$ 3 m room.
The reverberation time is chosen randomly in the 100 ms to 400 ms interval. 
The microphone spacing is 2 cm and the arrangement of sources is selected at random. All the models are trained for 20 epochs, and 100,000 mixtures are generated in each epoch by convolving the sources with the RIRs randomly selected from the training set. Another 800 and 1,600 mixtures are generated with unseen sources and RIRs for validation and test, respectively. 
All the signals are sampled at 16 kHz. The STFT frame size is 4096 with half overlap.
To eliminate scaling ambiguity of the separated signals, the estimated demixing matrix is rescaled based on the minimal distortion principle \cite{mini_distor}.
The iteration number of the demixing matrix updates is set to $L=5$. 

Besides the conventional methods, another two SOTA DNN-supported BSS algorithms are adopted for comparison (independent deeply learned matrix analysis (IDLMA) \cite{IDLMA} and the Kang method in \cite{DIVA_KANG}). To make a fair comparison, IDLMA and the Kang method use the same DCCRN architecture to directly learn their target from mixtures.
The performances of all the methods are evaluated by the SIR and SDR calculated by the BSS toolbox \cite{bss_measure}, and also word error rate (WER) and character error rate (CER) of an ASR system trained using the WSJ0 dataset \cite{wsj0} with the ESPnet framework \cite{ESPNET}.

\begin{figure}[!htbp]
	\centering 
	\includegraphics[width=0.498\textwidth]{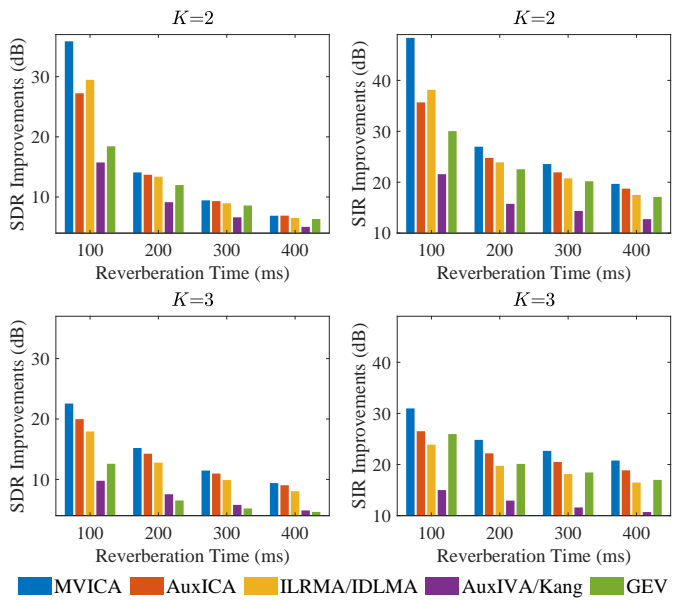}
	\caption{Oracle performances of different methods under various conditions.}
\end{figure}
\subsection{Results and Discussion}
\begin{figure}[!htbp]
	\centering 
	\includegraphics[width=0.495\textwidth]{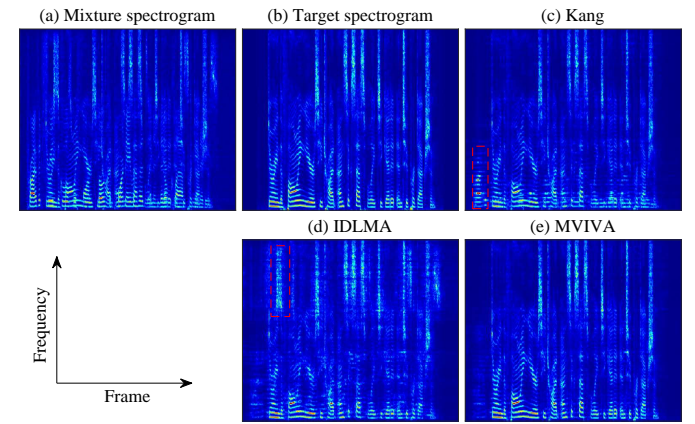}
	\caption{The spectrograms of a 2-source mixture (a), and the sources estimated by the Kang method (c), IDLMA (d) and the DNN-supported MVICA algorithm (e), where the red box marks the residual interference components.}
	\label{result_example}
\end{figure}

We begin by comparing the oracle separation performances that can be potentially achieved by different methods.
More specifically, the target spectrogram and frame-level variance of each source are directly given to AuxICA/ILRMA/IDLMA and the AuxIVA/Kang method, respectively. 
The proposed MVICA algorithm and the GEV beamformer (with blind analytical normalization \cite{GEV}) are also applied with the ideal interference covariance matrix.
As Fig. 1 shows, MVICA has the best oracle performance in terms of SDR and SIR, demonstrating its superiority over the existing BSS algorithms.
In addition, MVICA yields obviously superior performance over the GEV beamformer especially in terms of SDR scores, although they both aim to maximize the output SIR. 
Table 1 further shows the practical performances on the test set. The DNN-supported MVICA algorithm significantly outperforms the other methods with respect to SIR, SDR, WER and CER. The effectiveness of DNN-based interference covariance matrix estimation for MVICA is thus confirmed.
It is worth mentioning that, while MVICA is derived to maximize SIR, the other metrics have seen significant gains as well.
As an example, Fig. \ref{result_example} displays the source spectrograms predicted by different methods. It can be clearly seen that MVICA can recover the source signal with less residual interference.

\section{Conclusion}
In this letter, a new determined BSS algorithm framework called MVICA is mathematically derived towards maximizing the output SIR, where the optimal weighted covariance matrix is proved to be the interference covariance matrix. MVICA avoids the difficulty of source modeling and achieves the SOTA BSS performances. It is also superior to the MVDR and GEV beamforming techniques because it avoids the extra estimation of RTF and introduces fewer signal distortions.
In the future work, we intend to investigate more efficient ways to estimate the interference covariance matrix and apply the proposed methods to the noisy environments.

\end{document}